\documentclass[prb,preprint]{revtex4-1}

\usepackage{amsmath}  
\usepackage{amsfonts} 
\usepackage{graphicx} 
\usepackage{color,soul}
\usepackage{multirow}
\usepackage{array}
\newcolumntype{C}{>{\centering\arraybackslash}p{3.5em}}

\begin{document}

\preprint{}

\title{Index matching computerized tomography}
\author{Vincent Daley}
\affiliation
{Physical Sciences Department, Thompson Rivers University}
\author{Owen Paetkau}
\affiliation
{Department of Physics and Astronomy, University of Calgary}
\author{Mark Paetkau}
\affiliation
{Physical Sciences Department, Thompson Rivers University}

\begin{abstract}
Computerized tomography (CT) has been used for decades by medical professionals to detect and diagnose injuries and ailments. CT scanners are based on interesting physics, but due to their bulk, cost, and safety, hands on experience with a medical CT scanner is unrealistic for undergraduate students. Therefore, operationally similar, yet small, safe, and inexpensive CT scanners are desirable teaching tools. This project details the development of a novel model CT scanning apparatus. The experimental setup presented utilizes visible light, has short data acquisition time, and operates on the same physics as its X-ray counterpart. The apparatus employs a laser and photodiode to image a transparent material, while avoiding loss of transmitted intensity through index of refraction matching. A simple back-projection algorithm results in a 2D cross section of the scan object. We found we could collect data and reliably image samples in 15 minutes.

\end{abstract}

\maketitle

\section{\label{Intro}Introduction}

Computerized tomography (CT) imaging techniques have become common place in many fields, especially in medical imaging and diagnosis. The implementation of CT imaging has allowed for multislice X-rays to produce a 3D image of the relative densities within an X-ray transparent object. Data acquisition for a CT image begins by collecting X-ray projection data across a single 2D slice of the object of interest. A simple or filtered back-projection algorithm may then be employed to recreate this slice. Translating the object through an X-ray field will allow for several slices to be combined to produce a 3D image.\cite{ref1,ref2} Understanding the method of both data acquisition and reconstruction algorithms is important to students intent on working with CT imaging systems.

CT imaging systems, which are specified for clinical use, have high clinical demand and may not be available for teaching purposes. Simple and safe model CT imaging systems are valuable teaching tools. Mylott \emph{et al.} produced a model system employing a laser, photogate, and floral foam cylinders.\cite{ref3} This system was able to create a 2D projection of the cylinder outlines with an imaging time of around 20 minutes. Several other studies have produced experimental setups employing CT scanning techniques employing light from both infrared and visible spectra.\cite{ref4, ref5, ref6, ref7} Another study completed by Paetkau \emph{et al.} employed a $^{90}$Sr beta source and Geiger counter to produce and detect beta rays through a floral foam object with offset cutout.\cite{ref8} The beta-ray system reproduced the physics of an X-ray system and was able to accurately image a 1-cm wide, bean-shaped hole within the foam.  The drawback of the beta-ray system was the imaging time of 1.25 hours. 

The ideal setup employs the use of a laser and photodetector for fast scanning time, while imaging a transparent material and avoiding reflection and refraction. Glass is a transparent material in the visible spectrum but, in air, a glass rod redirects light by reflection and refraction. However, a glass rod immersed in a liquid of identical index of refraction will eliminate reflection and refraction, causing the glass rod to ``disappear.''\cite{ref9} This effect may be leveraged to employ the fast scanning laser and photogate setup shown by Mylott \emph{et al.}\cite{ref3} Furthermore, the use of colored glass will cause absorption, 
%
dependent on the color of glass, which simulates material with spatially variant density.\cite{ref10} 
%
This paper presents a novel model CT imaging setup, which is able to simulate an X-ray transparent system. Our goal in designing this setup was to provide a short imaging time demonstration of the CT scan experimental technique and reconstruction methods for use in the undergraduate physics instructional laboratory.

\section{\label{Theory}Theory}

X-rays used in clinical CT imaging systems are neither reflected or refracted from the material being imaged, so the imaging is based on the relative attenuation of tissues. This attenuation is reported as the mass attenuation coefficient $\mu$, which is dependent on the density and effective atomic number of the imaged material.\cite{ref1} The mass attenuation coefficient is energy dependent and a known quantity for most organs. As X-rays pass through a single organ, the attenuation is exponential in form.  In our system, visible light replaces X-rays. Below, we show that by matching refractive indices, reflections and refractions are eliminated.  Furthermore materials may have different attenuation coefficients, thus with the correct choice of materials, visible light can replace X-rays in a model CT imaging setup.  

Light incident upon a material boundary is refracted according to the angle of incidence $\theta_i$ and the refractive indices $n_1$ and $n_2$ of the two adjacent media. The relation is given by Snell's Law:

\begin{equation}
    n_1 \sin\theta_i = n_2 \sin\theta_t.
    \label{eq:snell}
\end{equation}

When the refractive indices are matched (i.e., equal), Eq.~(\ref{eq:snell}) shows the light's incident angle $\theta_i$ and the transmitted angle $\theta_t$ are equal.  Thus, light will travel in a straight line through the  interface.  

Light incident on a boundary is also reflected and follows the reflection law: $\theta_i$=$\theta_r$.  The amount of light reflected is governed by the reflection coefficient $r$, which is the ratio of the reflected electric field amplitude to the incident.  Reflection coefficients are usually written for \emph{p}-polarization (in plane of incidence) and \emph{s}-polarization (perpendicular to plane of incidence).  Using this general approach, the reflection coefficients can be written as Fresnel's sine and tangent laws:
\begin{equation}
r_p = \frac{\tan(\theta_i-\theta_t)}{\tan(\theta_i+\theta_r)}
 \label{eq:reflectionp}
\end{equation}
\begin{equation}
r_s = -\frac{\sin(\theta_i-\theta_t)}{\sin(\theta_i+\theta_r)}.
 \label{eq:reflections}
\end{equation}
When indices are matched, the incident and transmitted angles are equal and we see $r_p=r_s=0$, i.e., no light is reflected. 

When light is incident on two different materials with matched indices of refraction, the light is not reflected and is not refracted, and so propagates along a straight line through the materials.  However, while two materials may have matched indices, they may absorb the light differently.  The absorption is modeled as an exponential decay:
\begin{equation}
    I = I_0 e^{-\mu \Delta x}
    \label{eq:absorption}
\end{equation}
where $I$ is the final light intensity, $I_0$ is the initial light intensity, $\Delta x$ is the thickness of the absorbing material, and $\mu$ is the attenuation coefficient of the material.

The simple mechanism of absorption becomes more complex as light passes through many different materials, each contributing intensity loss.  In this case, the intensity expression becomes an exponential sum of the coefficients weighted by the distance through each material
\begin{equation}
    I = I_o e^{-(\mu_1 \Delta x_1 + \mu_2 \Delta x_2 + ... + \mu_i \Delta x_i)}.
    \label{eq:5}
\end{equation}
This extension of the attenuation expression is the basis for clinical CT reconstruction, where the goal is to determine the unknown attenuation coefficients at each pixel throughout the image.  From a medical perspective, different internal structures have different densities, hence different attenuation coefficients.  Our index-matched visible-light system uses Pyrex tubes immersed in mineral oil to provide index-matching. By varying the color of these Pyrex tubes, a spatial variation of the attenuation coefficient is produced.\cite{ref10} Since mineral oil and Pyrex have refractive indices near 1.48 $\pm$ 0.01,  reflections and refractions are eliminated in the optical set-up leaving the attenuation coefficient as the dominant signal reduction effect.\cite{ref9}
%
Th attenuation effect may be leveraged to produce the CT projection data required for image reconstruction. The Pyrex vials are hollow and cylindrical producing a difficult, non-uniform thickness material to image. The thickness of the Pyrex affects the total attenuation and therefore the reconstruction data as shown in Eq. \ref{eq:5}. An expression to describe the expected signal after the laser has passed through the Pyrex vials may be used to determine the attenuation coefficients from the measured data. The expected signal expressing, $f(x)$, becomes complex as the laser is translated across the face of the cylinder due to the varying thicknesses. The location of $x=0$ is taken to be the center of the cylinder, with a piece-wise function describing the attenuation as seen below:
%

\begin{equation}
f(x) = \left\{
        \begin{array}{ll}
            e^{-2\mu\left[\sqrt{r_0^2-x^2}- \sqrt{r_i^2-x^2}\,\right]}  & \quad 0\leq x < r_i , \\
            e^{-2\mu\sqrt{r_0^2-x^2}} & \quad r_i \leq x < r_o , \\
            0 & \quad x \geq r_o .
        \end{array}
    \right.
    \label{eq:attenuation}
\end{equation}

%

In Eq.~(\ref{eq:attenuation}), $\mu$ is the attenuation coefficient, $r_o$ is the outer radius of the Pyrex tube, $r_i$ is the inner radius of the Pyrex tube and $x$ is the distance from the center of the tube. The piece-wise function describes a region through which the laser passes through the entire width of both sides, through the thick edges and once the position is outside the outer radius using the equation of a circle as the basis for derivation. The radii can be measured directly, and therefore the attenuation coefficient can be determined from a best fit of the data to this function.

\section{\label{Algorithm}Back-Projection Algorithm}

A simplified back-projection, described in great detail by Delaney and Rodriguez,\cite{ref4} is employed in our experiment, allowing the projection data collected from successive rotations and translations to be formed into a CT image. The algorithm is a straightforward method of generating an image based on absorption measurements. Filtered back-projection can also be applied, offering improved image quality.\cite{ref1,ref2,ref4,ref8}

\begin{figure}[h]
\centering
 \includegraphics[width =150mm]{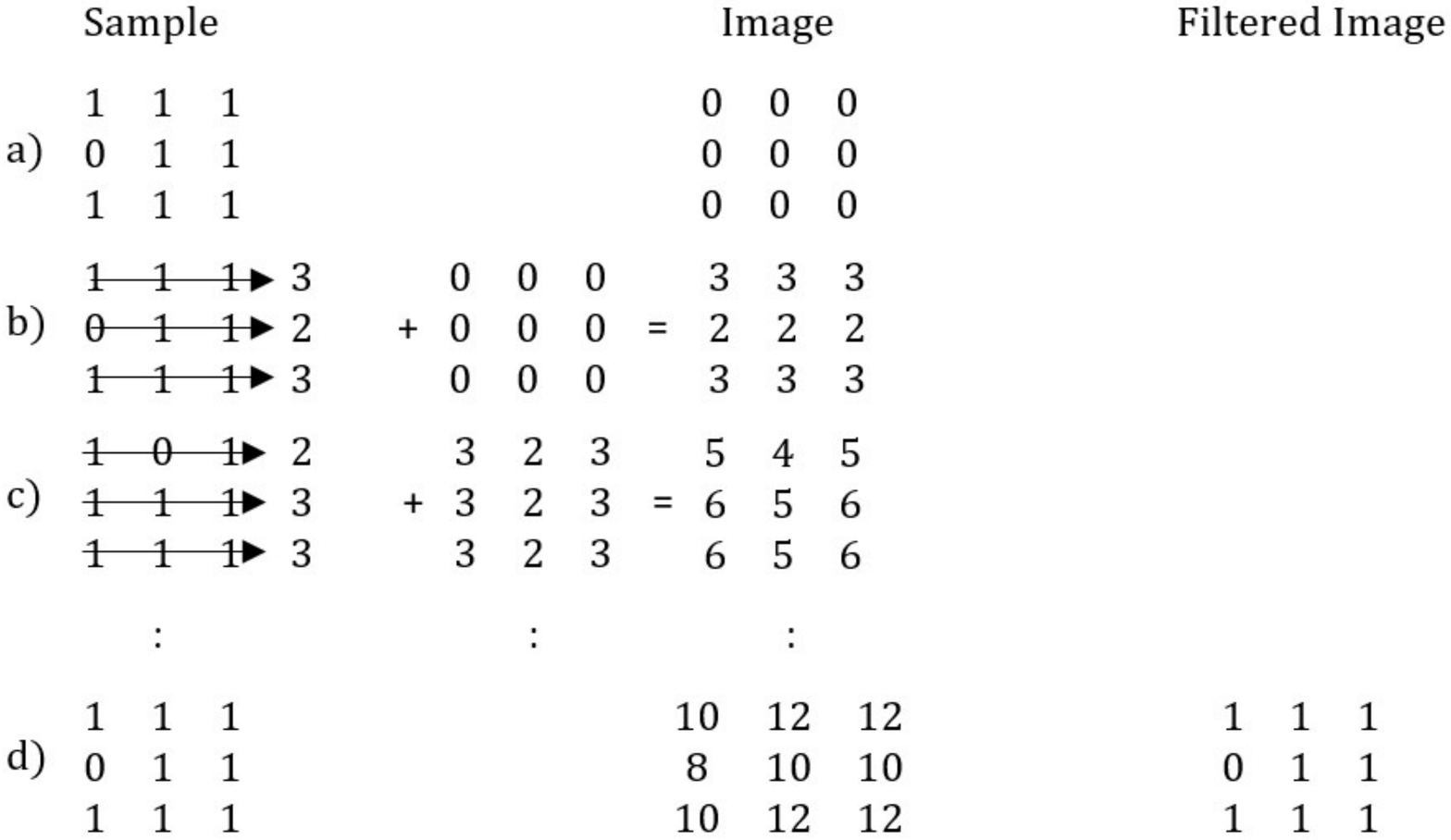}
    \caption{
        Explanation of the back-projection algorithm.  We start with (a) a sample which has a hole (the single 0) and a blank image (all zeroes). (b) shows the result of a scan along the three rows which evolves the image. (c) both the sample and the image are rotated, the scan along the rows is again added to the image.  For (d) we repeat the procedure for two more rotations and return the sample to its original orientation.  By filtering the image, the sample is duplicated. 
    }
  \label{fig:algorithm}
\end{figure}

In the Delaney and Rodriguez paper,\textsuperscript{4} a graphical explanation is used, so we will look at a simple numerical explanation.  We start with a sample represented by a 3$\times$3 matrix, where the numbers represent absorption of the pixel; as seen in Fig.~\ref{fig:algorithm}, `0' is a hole or air gap, `1' is a region of large absorption (or material).  
%
On the right is the back projection image; as we have no scan data at this point, the image consists of zeroes.

In CT imaging we cannot probe each element individually, only groups of elements, and for this example, probing means to sum the absorption coefficients.  We look at each row and see that a scan to the right gives the column vector (3 2 3), as shown in row b of Fig. 1.  This is our first information on the image, so we add this information to the image, and the back-projection starts to take form.

Now we rotate the sample (and the back-projection image) 90 degrees clockwise (Fig. 1 row c), and again we add the result to the reconstructed image.  If we repeat the procedure two more times, then return to the original orientation, the image now begins to reflect the nature of the sample (at least in the sense that the minimum values are at the same location).  If we go one step further and know we have a binary system (only `1's and `0's), then we can filter the image, resulting in a match between the image and the sample. In an actual one of our scans, there are many more rows and columns 
(40) and many more rotational projections 
%
(400), which increases the resolution of the image.

Intensity is modeled as exponential decay [Eq.~(\ref{eq:absorption})], so the absorption can be calculated as:\cite{ref4}
\begin{equation}
    A(x, y) = \ln|I_0| - \ln|I(s,\theta)| 
    \label{Eq2}
\end{equation}
$I_0$ is the intensity from the source without any interfering medium (maximum or average), and $I(s,\theta)$ is the intensity transmitted through the material at translation $s$ and rotation $\theta$. $A$ is actually $\mu \Delta x$, but since the $\Delta x$'s are the about the same, we call this the absorption.  When radiation is directed through the material $I(s,\theta) < I_0$, the absorption $A$ is positive. Regions of little absorption mean $I_0 \approx I$ and result in $A\approx0$.  When plotted as a grayscale image, the image appears as an classic X-ray image: regions of low absorption are black and regions of high absorption are white. The back-projection algorithm described above was coded in the R programming language using the pseudo code below to produce the desired results.

Since the image is made of discrete points $P_{ij}=(x, y)$, Delaney \emph{et al.} note, for a sample translated $s$ and rotated $\theta$, the image pixels fall on the line
within a distance $D$ of the following line:
\begin{equation}
    y\cos\theta -x\sin\theta-s=0
    \label{eq:inequalityA}
\end{equation}
The pixels are square, so to determine the pixels that the line intersects, an inequality is used:
\begin{equation}
    y \cos\theta - x \sin\theta - s \leq D 
    \label{eq:inequalityB}
\end{equation}    
where $D$ is taken as half the translation increment.  

The process of constructing an image of the sample proceeds as follows.  The absorption at $(x,y)$ is initially set to zero.  Then, for each translation and rotation $(s,\theta)$, if the inequality [Eq.~(\ref{eq:inequalityB})] is satisfied, then the absorption $A(x,y)$ is calculated and incremented.  The process is repeated for all pixels $(x, y)$ in the image.
\begin{verbatim}
Initialize the image array, A (81x81), to zero
Initialize spatial coordinates array X and Y 
Initialize Angle_array (0 to 359.1 in steps of 0.9)
Initialize the translation array, S (-2 to 2 in steps of 0.1)
D=0.05
Read data into array N, sizeof(S) rows x sizeof(Array_angle) columns
Use average of unobstructed transmission data to determine No
for all image points (i,j)
  for all data points(k,l)
    if abs(Y[j]cos(Angle_array[l])-X[i]sin(Angle_array[l])-S[k])<=D then
          increment A[i,j] by ln|No| - ln|N[k,l]|
Plot the array, A, as an image using coordinates X and Y
Apply filter to plot a binary image as needed
\end{verbatim}

\section{\label{Method}Method}

Data collection was automated using an Arduino Uno microcontroller board, stepper motors, laser pointer, and photodiode. The photodiode used was a Hinds Instruments DET-90 silicon photodiode detector. The motor drivers were SparkFun ROB-12859 Big Easy boards. For the linear rail and stepper, OpenBuilds V-Slot Mini V Linear Actuator Bundle with a NEMA 17 stepper were used and the rotary stage was comprised of a Kurokesu RSA1 also paired with a NEMA 17 stepper. The laser was battery powered pen laser-pointer with a 5 mW maximum output and 405 nm wavelength. 

Samples were colored Pyrex glassblowing tubes.  The tubes were bathed in heavy mineral oil (a mild laxative found at most drug stores). Corn or vegetable oil also would work, but mineral oil has the advantage of being colorless. The index of refraction of the mineral oil was measured using a hollow prism and the minimum deviation method.\cite{ref11} The index of refraction of the mineral oil was measured to be 1.489 $\pm$ 0.001 for the 410 nm line of a hydrogen source.

Sample tubes were mounted on the rotating stage with the bottom part of the tube immersed in the oil.  The samples of interest were a blue Pyrex tube of inner diameter 6.90 $\pm$ 0.05 mm and outer diameter of 10.25 $\pm$ 0.05 mm and a dark green Pyrex tube of inner diameter 6.80 $\pm$ 0.05 mm and outer diameter of 10.05 $\pm$ 0.05 mm.

\begin{figure}[ht]
   \centering
    \includegraphics[width =150mm]{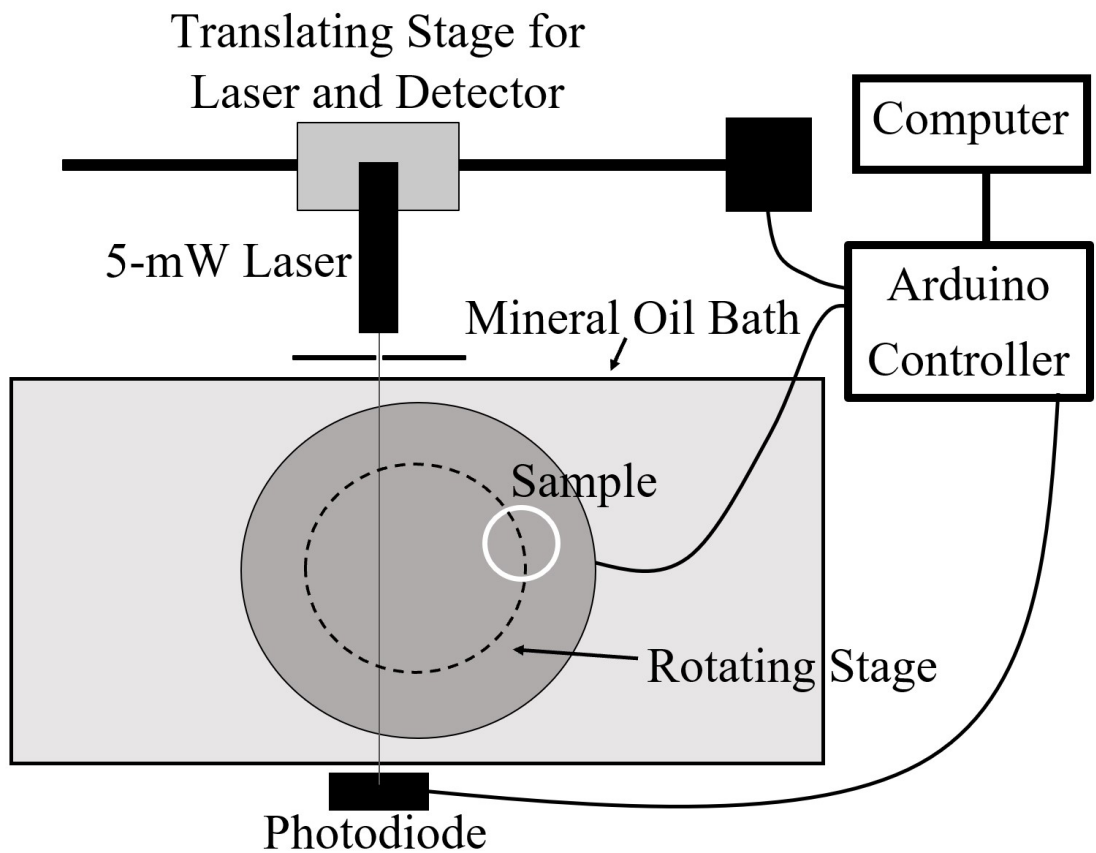}
       \caption{
Diagram of CT Scanner used in this experiment }
    \label{fig:Diagram}
\end{figure}

Figure~\ref{fig:Diagram} illustrates the model laser CT scan setup we used. The Arduino board controlled the rotating and translating stages. The rotating stage remained stationary and was elevated above the oil bath. The tubes were suspended from the rotating stage and dipped into the oil. The laser and detector were mounted to the lateral stage, allowing them to travel from the right edge of the scan area to the left. A 1-mm collimating slit was placed between the laser and the oil bath. The laser was left on for the duration of a scan while the microcontroller measured the voltage across the photodiode. A serial monitor captured the readings at each interval of the scan. The scans appearing in this paper were imaged over a physical distance of 42 mm. The adjustable parameters of lateral increment, rotational increment, and measurement time  were set to 1 mm, 0.9 degrees and 10 ms, respectively. The movement sequence of the apparatus was as follows: set rotation stage to zero, read data, increment rotation by 0.9 deg, read data, repeat until 359.1 degrees is reached. 
%
After the rotation sequence, a linear step occurs and the rotation process is repeated. Previous setups use 5 or 10 degree increments with averaging, but data collection time was minimized and reconstructed images were of highest quality with more angular steps and less averaging. Typical data array sizes obtained were 42$\times$400. Image reconstruction was achieved using a back-projection algorithm written in R, version 3.5.3.\cite{ref1,ref3,ref4,ref8} 

Image analysis was used to measure the diameters, thicknesses, and other geometric characteristics of the tubes. Lower and upper window limits were provided to isolate specific relative absorption levels. This provided a mask on which the inner and outer diameters might be measured. The inner diameter was determined from the area enclosed by the mask and the outer was determined from the limits of the mask. These circles were plotted around the centroid determined by a center-of-mass calculation for each tube. These centroids could be used to verify distance metrics as compared to the experimental setup. 

To test the quality of the index matching, a linear scan was made of three tubes in air and in oil.  In this case, the laser was translated across the width of the tube (no rotation), while the microcontroller measured the voltage from the photo-diode.  The linear step size in these measurements was about 0.03mm.  

\section{\label{Results}Results}
\subsection{\label{Index Matching}Index Matching}

\begin{figure}[ht]
   \centering
    \includegraphics[width =150mm]{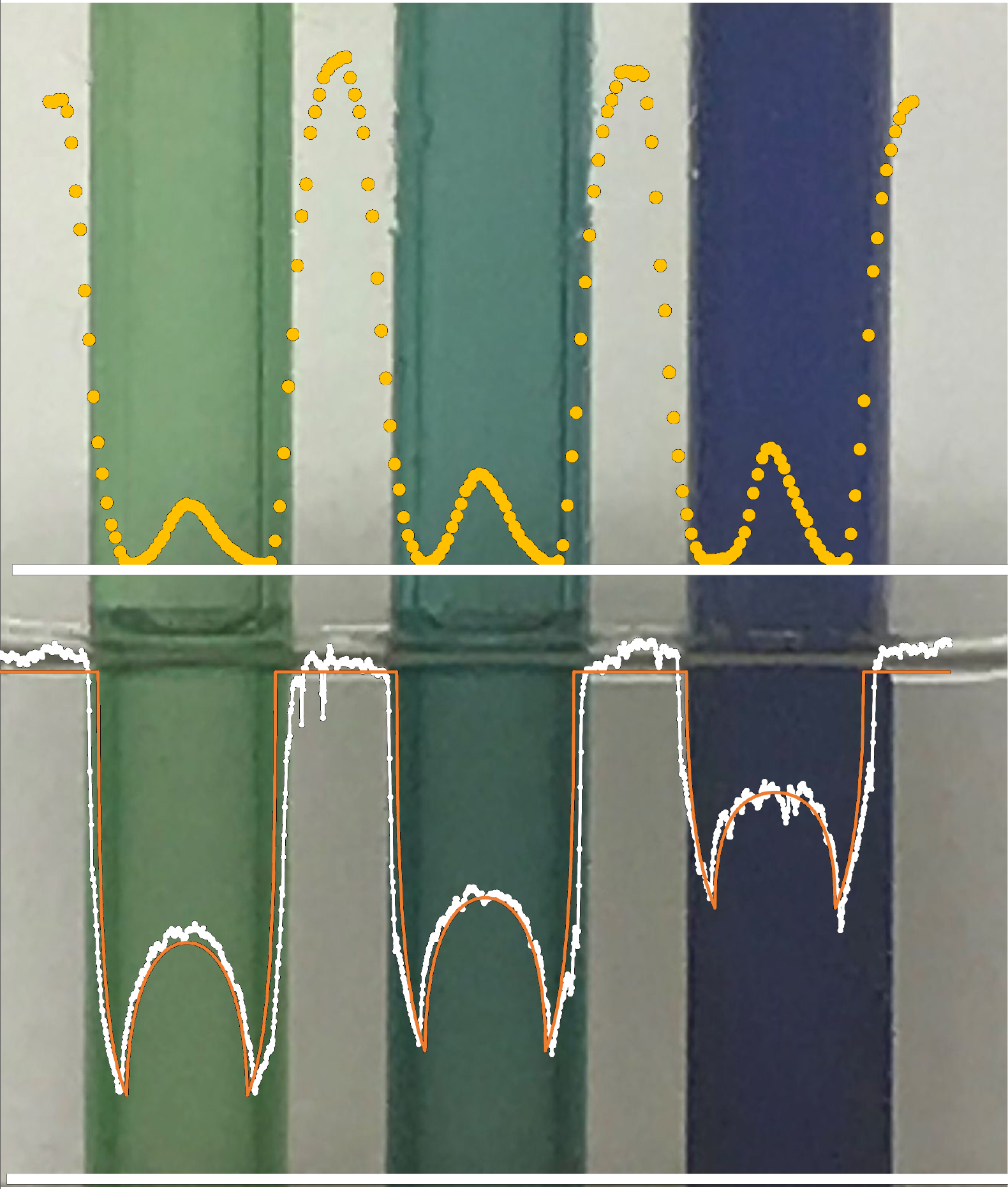}
       \caption{Light green, dark green and blue Pyrex tubes submerged in mineral oil. Due to index matching, the edges of the tubes disappear and only the coloring contaminants remain visible. Absorption curves taken in air and submersed in oil using a 405 nm wavelength laser and are superposed on the image of the tubes. The plots have intensity on the \emph{y}-axis and displacement on the \emph{x}-axis. The orange line is a theoretical curve using measured absorption coefficients of the Pyrex tubes. The thick white lines represents zero intensity for the data in air and in oil. }
    \label{fig:AbsorptionAndIndexMatch}
\end{figure}

Figure~\ref{fig:AbsorptionAndIndexMatch} is a twist on the familiar ``vanishing stir rod'' experiment.\cite{ref9} The top half of the Pyrex tubes are in air and the bottom halves are in mineral oil.  The oil surface can be seen at the midway point of the image, just below the thick white line.   The yellow lines in the top half of the image is the voltage of the photodiode (intensity) as the laser is scanned across the three tubes.  The data have been normalized.  The white line represents zero.  As the laser encounters a tube, the intensity goes to zero as the light is reflected/refracted away from the detector.  At the center of the tube (normal incidence), some light is detected, but clearly much of it is still reflected away.  

The bottom halves of the tubes are immersed in mineral oil and the intensity plot is very different.  Again the data are normalized and the broad white line is zero.  In this case, as the laser reaches the edge of a tube, the intensity drops, but not to zero. As the laser reaches the inner edge of the tube, the intensity is minimized as this is the thickest part of the glass.  The intensity recovers as the laser traverses the inner edge, reaching a local maximum (minimum thickness of colored glass) at the midpoint. The data are symmetric about the tube.  We see that the different colors of glass have different absorption coefficients.  This behavior can be modeled using 
%
Eq.~(\ref{eq:attenuation}).  The least-squares best fit is shown by the orange line.  The fitting is based only on attenuation properties of the glass, so we see the index matching has done an excellent job of removing reflection/refraction effects. The attenuation coefficients were measured to be 2.30 $\pm$ 0.03 cm$^{-1}$, 1.78 $\pm$ 0.03 cm$^{-1}$ and 0.82 $\pm$ 0.03 cm$^{-1}$ for the light green, dark green, and blue Pyrex tubes respectively. 

\subsection{\label{CT Images}CT Images}

\begin{figure}[ht]
  \centering
   \includegraphics[width =150mm]{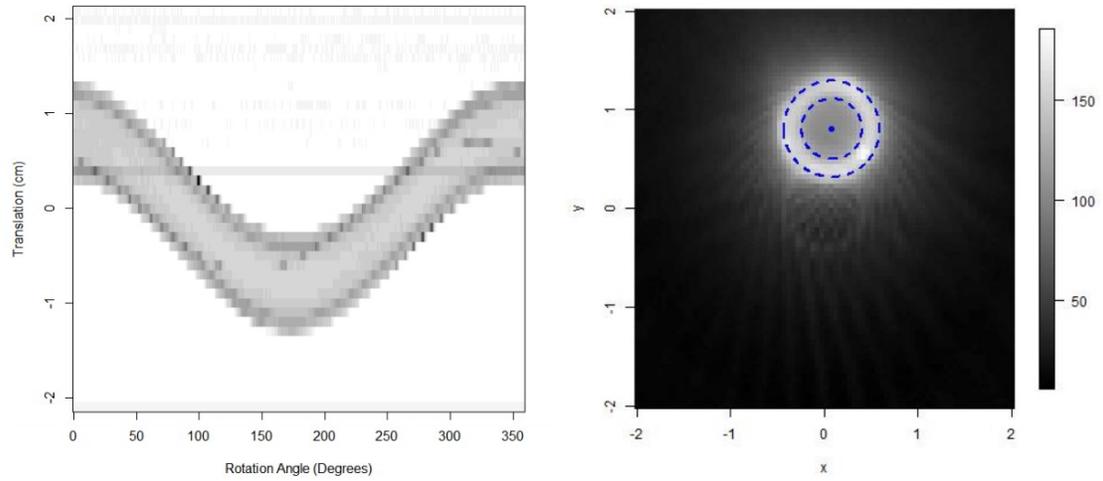}
      \caption{The left image displays the sinogram while the right image displays the reconstructed image. The tube was imaged with a purple laser using 1 mm, 0.9 degree steps. Dashed lines indicate the detected boundaries of the tube and the point indicates the centroid.}
    \label{fig:SingleVials}
\end{figure}

A CT scan of the blue Pyrex tube using the purple 405 nm laser with 1 mm translation and 0.9 degree rotation steps was completed. Each scan produces two useful images, as shown in Fig.~\ref{fig:SingleVials}, the sinogram of the data and the reconstructed image. The sinogram displays the raw collected data for each translation and rotation. If we follow the sinogram from left to right we see where the tube is at any degree of rotation. We see that at zero degrees, the center of the sinusoid is at the same position on the \emph{y}-axis as the center of the tube. So we can see that by following the center of the tube as it rotates through 360 degrees, the trajectory is sinusoidal. Additionally, the sinogram displays raw data so regions of high absorption (i.e., low detected intensity) appear dark. Conversely, in the CT images such regions have high intensity, white coloring, consistent with their higher attenuation coefficients. The glass tubes have much higher attenuation than the surrounding oil, thus they appear white. Inner and outer diameters of both tubes were measured from the reconstructed images using 
%
windowing techniques. The window levels were selected using a value relative to maximum pixel value observed of the vials. They were adjusted manually using trial and error until they fit on the images and the diameters were subsequently measured. The blue tube had a measured inner diameter of 6.4 $\pm$ 0.2 mm and an outer diameter of 10.0 $\pm$ 0.2 mm as indicated by the dashed blue lines. The centroid lays 8.1 $\pm$ 0.2 mm away from the scan center.

Further imaging, a single dark green Pyrex tube and the blue and dark green Pyrex tubes in combination, is shown in Fig.~\ref{fig:TwoVials}. The respective sinograms are displayed to the left of the images. In the single dark green Pyrex tube scan, the tube had a measured inner diameter of 6.2 $\pm$ 0.2 mm and an outer diameter of 9.8 $\pm$ 0.2 mm with the centroid lying 12.7 $\pm$ 0.2 mm away from the scan center. Meanwhile in the two tube scan, the dark green tube offered higher contrast, but both tubes were resolvable. Notice that the green tube, which has a greater attenuation coefficient than the blue tube, is brightest. The dark green Pyrex tube had an inner diameter of 6.2 $\pm$ 0.2 mm and outer diameter of 10.0 $\pm$ 0.2 mm. The blue Pyrex tube measured to have an inner diameter of 6.2 $\pm$ 0.2 mm and an outer diameter of 10.2 $\pm$ 0.2 mm. The distance between the blue and green Pyrex tube centroids was measured to be 20.2 $\pm$ 0.2 mm.

In all three CT images the inner disks of the tubes appear brighter than outside the tubes, despite both regions being mineral oil and having the same attenuation coefficient. In Fig.~\ref{fig:SingleVials} we also observe a ring, with similar diameter to the tube, around the origin of the plot area, this seems to correlated with the horizontal line in the sinogram, so a slight glitch at the linear position. 

The imaging time for these scans was less than 15 minutes, in which time 17,200 data points were collected. A further minute was then required for image reconstruction when run on a Microsoft Surface Pro (Intel i5-7300 processor, 8 GB RAM).

\section{\label{Discussion}Discussion}

Teaching students the basics of computed tomography scans and the related reconstruction algorithms is a valuable component of an undergraduate education.\cite{ref2} CT techniques are employed in many scientific and medical fields and access to X-ray CT apparatus is not often feasible. The availability of materials, low scan times and semi-transparent samples are the limitations of model CT scanners. Previous work accomplished similar results using either a laser or a beta-ray source, but had failed to combine low scan time, transparent object imaging, and low-cost as we have here.\cite{ref3,ref4,ref8} Our setup is easily attainable, equipment costs amounted to roughly \$500 CAD. The index matching to produce a semi-transparent material has made for an effective analog for varied attenuation coefficients. Figures 4 and 5 provide sinograms and reconstructed CT scans of Pyrex tubes. In each case, both the inner and outer radii of the semi-transparent material were resolvable. These results were achieved while requiring about 15 minutes for data acquisition and image reconstruction. 

\begin{figure}[ht]
    \centering
     \includegraphics[width =150mm]{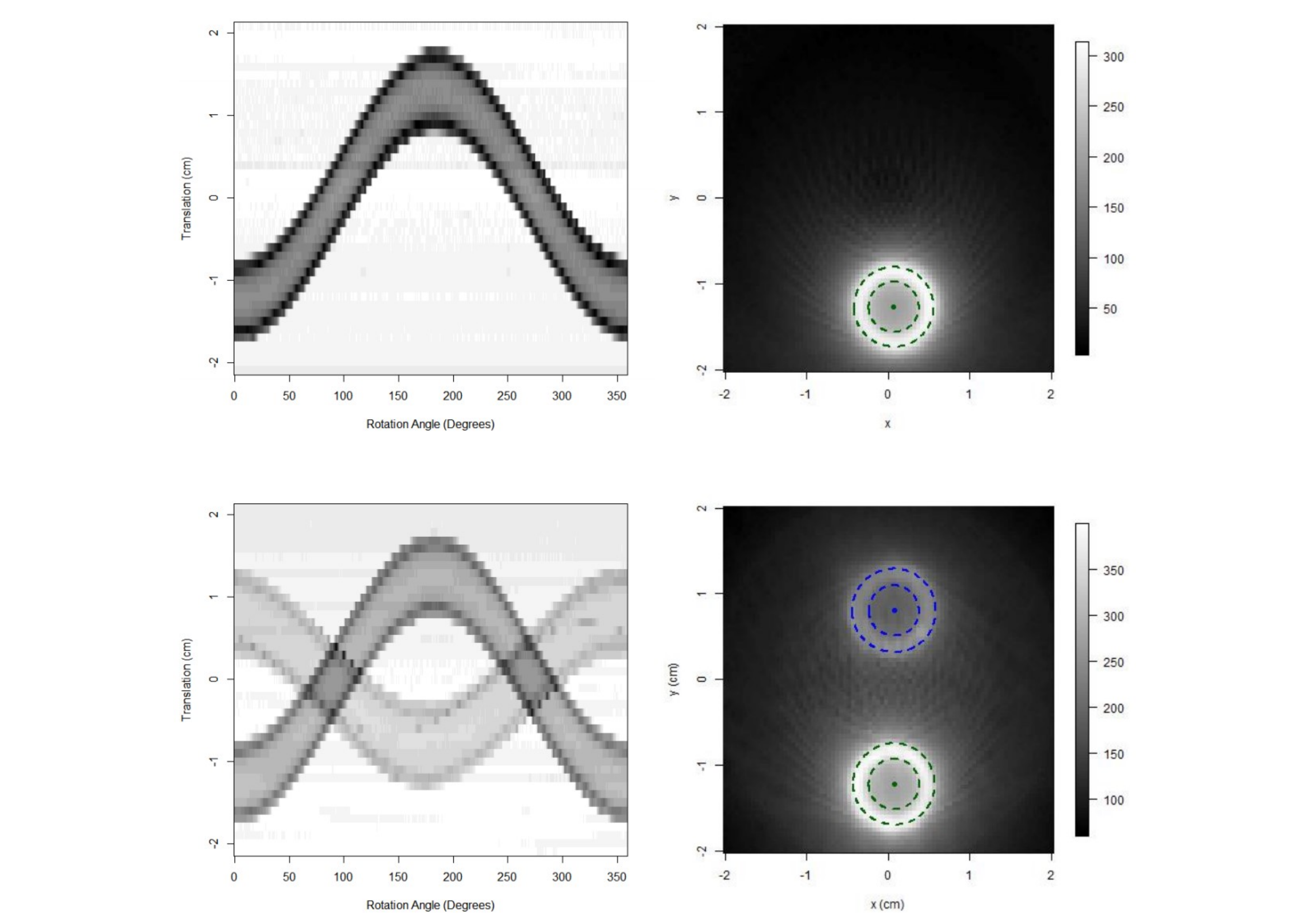}
      \caption{Dark green and blue Pyrex tubes imaged with 1 mm translations and 0.9 degree rotations. Left shows the sinogram  while right indicates detected inner and outer radii on the reconstructed images.}
    \label{fig:TwoVials}
\end{figure}

The accuracy of the CT images shown in Figs.~\ref{fig:SingleVials} and~\ref{fig:TwoVials} was first evaluated using several geometric measurement. The outer and inner radii of each Pyrex tube were measured first using vernier calipers to provide a reference value. The same radii were then measured in the one tube and two tube scans to produce quantitative values to compare to evaluate the accuracy of the model CT scans with results displayed in millimeters in Table \ref{tab:radii}. The model CT platform was able to measure the inner radius to within 0.7 mm for the one and two tube scans. Additionally, the outer radius was measured to within 0.25 mm of the reference value. Accurate measurements using this model CT platform were possible within the uncertainties provided. 

The distance between the two tubes in the double scan center was measured from Fig.~\ref{fig:TwoVials} to be 19.8 $\pm$ 0.2 mm   close to the 20.0 $\pm$ 0.5 mm physical separation between the tubes 

Image quality was impacted by several factors. Reflection and refraction by the glass were reduced significantly, but we cannot confidently state they were eliminated. Variance in production of the Pyrex, such as imperfections, could affect the refractive index of the glass. Interestingly, scans attempted with a red laser were more susceptible to adverse effects due to refraction. The index of the oil at red wavelengths was measured to be 1.475 $\pm$ 0.001, so the Pyrex may have had a higher index of refraction than the reference value of 1.47.\cite{ref9} Furthermore there was evidence the vessel holding the mineral oil contributed to imperfect images.  The vessel was a plastic box and there is evidence the sides of the box are slightly bowed, leading to reflections.  Future improvements will use a container of flat Pyrex glass (likely biological specimen slides).  Fortunately, it was not imperative that the refractive indices be matched exactly provided the refraction did not obscure the change in absorption through different points in the tubes. It was also important the levels recorded by the photodiode be significantly different in material and in air. The plots seen in Fig.~\ref{fig:AbsorptionAndIndexMatch} indicate the difference in attenuation between air and oil scans. 

Ensuring the origin of the reconstruction corresponded to the center of rotation of the sample was also important to maximize image quality. If the center was off by as little as 1 mm, image quality was severely deteriorated. The sinograms seen in Figs.~\ref{fig:SingleVials} and~\ref{fig:TwoVials} were useful to verify centering. We ensured the peaks and troughs of the sinogram were equally spaced from the zero line to confirm the centering. If off-center, the scan could be re-centered in image reconstruction although it was preferred the scan center was known during data collection. Confirming the image was centered prior to image reconstruction greatly improved image quality. 

The sophistication of our image reconstruction algorithm is another major consideration of image quality. We employed simple back-projection, but it is known that filtered back-projection offers superior image reconstruction.\cite{ref1,ref4,ref8} Use of a more sophisticated reconstruction algorithm may remove the elevated intensities inside of the tubes noticed in the reconstructed images.  However, the implementation of filtered back-projection is beyond the scope of this paper.

In addition to providing an introduction to the physics of X-ray CT scanning in its current configuration, the apparatus is ideal for further student investigations.  The effects of laser wavelength and tubes colors may be investigated, or the effects of using a different oil.  The index of refraction might be subtly altered by mixing another fluid. More complex shapes composed of Pyrex could be scanned to determine the level of detail the scanner can resolve. One of the improved image reconstruction algorithms alluded to previously\cite{ref1,ref4,ref8}
%
 can be employed and compared to the results presented here. Finally, the data collection sequence and intervals are not necessarily optimal, time savings may remain available in data collection.

The results presented indicate our index-matching CT scanner is an excellent, low cost, fast, and safe apparatus for introducing students to the physics of CT scanners.
\section{\label{Conclusion}Conclusions}
We have assembled a novel laser-sourced CT scanning apparatus, which allows for fast scanning time while resolving the inner and outer boundaries of a transparent object. The system makes use of a laser pointer, photodiode detector, stepper motors, and colored Pyrex glass tubes. The equipment is available in most undergraduate physics departments, making the apparatus an ideal platform for university physics students to explore CT scanning. The device presents a powerful and multifaceted learning experience, allowing students to explore the absorption of radiation by different objects and optics as well as computer based data acquisition and analysis.
%
\section{\label{sec:level1}Acknowledgements}
We are very appreciative of the excellent work by Dave Pouw in setting up the stepper motors and controllers.

\begin{table}[ht]
    \caption{Measured Radii of Pyrex Tubes}
    \begin{ruledtabular}
    \begin{tabular}{l c c c c c c}

        & \multicolumn{2}{c}{Caliper radii (mm)} & \multicolumn{2}{c}{Single tube scan radii (mm)} & \multicolumn{2}{c}{Dual tube scan radii (mm)} \\
       Tube & \multicolumn{2}{c}{$\pm$ 0.05} & \multicolumn{2}{c}{$\pm$0.2} & \multicolumn{2}{c}{$\pm$ 0.2} 
      \\ \hline
        & Inner & Outer & Inner  & Outer & Inner & Outer \\
        Blue & 6.90 & 10.25 & 6.4  & 10.2 & 6.2 & 10.2 \\
        Dk Green & 6.80 & 10.05 & 6.2 & 9.8 & 6.2 & 10.0\\ 

    \end{tabular}
    \end{ruledtabular}
 \label{tab:radii}
\end{table}

\begin{thebibliography}{11}
\bibitem{ref1}
J. T. Bushberg, J. A. Seibert, A. M. Leidholdt, and J. M. Boone, \textit{The Essential Physics of Medical Imaging} (Lippincott Williams \& Wilkins, 2011), pp. 350-357.
\bibitem{ref2}J. Fagerstrom, ``Computed tomography, sinograms, and image reconstruction in the classroom," Phys. Educ. \textbf{55} (3), 1-6 (2020). 
\bibitem{ref3}E. Mylott, R. Klepetka, J. C. Dunlap, and R. Widenhorn, ``An easily assembled laboratory exercise in computed tomography," Eur. J. Phys. \textbf{32}, 1227-1235 (2011). 
\bibitem{ref4}C. Delaney and J. Rodriguez, ``A simple medical physics experiment based on a laser pointer," Am. J. Phys. \textbf{70}, 1068-1070 (2002). 
\bibitem{ref5}I. G. Darvey, ``A simple inexpensive procedure for illustrating some principles of tomography," Phys. Teach. \textbf{51} (5), 298-299 (2013). 
\bibitem{ref6}O. Zietz, E. Mylott, and R. Widenhorn, ``Infrared radiography: Modeling X-ray imaging without harmful radiation," Phys. Teach. \textbf{53} (1), 46-49 (2015). 
\bibitem{ref7}Y. De Deene, ``Optical CT scanning for experimental demonstration of medical x-ray CT and SPECT," Eur. J. Phys. \textbf{40} (2), 1-20 (2019). 
\bibitem{ref8}O. Paetkau, Z. Parsons, and M. Paetkau, ``Computerized tomography platform using beta rays," Am. J. Phys. \textbf{85}, 896-900 (2017).
\bibitem{ref9}R. Budwig, ``Refractive index matching methods for liquid flow investigations," Exp. Fluids \textbf{17} (5), 350-355 (1994). 
\bibitem{ref10}B. S. Bae and M. C. Weinberg, ``Optical absorption of copper phosphate glasses in the visible spectrum," J. Non-Cryst. Solids \textbf{168}, 223-231 (1994). 
\bibitem{ref11}R. Synowicki, G. K. Pribil, G. Cooney, C. M. Herzinger, S. E. Green, R. H. French, M. K. Yang, J. H. Burnett, and S. Kaplan, ``Fluid refractive index measurements using rough surface and prism minimum deviation techniques," J. Vac. Sci. Technol. B: Microelectronics and Nanometer Structures Processing, Measurement, and Phenomena \textbf{22} (6), 3450-3453 (2004).
\end{thebibliography}
\end{document}